# Towards Non-Contact Glucose Sensing in Aqueous Turbid Medium at ~1.1 Meters Distance

Daqing Piao*,[1] Senior Member, IEEE, John O'Hara,[1] Senior Member, IEEE, Satish Bukkapatnam,[2] and Sabit Ekin*,[1] Member, IEEE

[1]School of Electrical and Computer Engineering, Oklahoma State University, Stillwater, OK 74078-5032, USA.
[2]Department of Industrial and Systems Eng., Texas A&M University, College Station, TX 77843-3127, USA.

**Abstract:** This work demonstrates a non-contact diffuse reflectance approach with a working distance of ~1.1 meters for the potential of glucose sensing. Non-contact diffuse reflectance over 1.1-1.3 µm was developed according to a center-illumination-area-detection (CIAD) geometry. The modeled response of diffuse reflectance in the CIAD geometry was examined with phantoms by altering independently the size of the collection geometry and the reduced scattering and absorption properties of the medium. When applied to aqueous turbid medium containing glucose control solutions with the cumulative volume varying over three orders of magnitude, a linear relationship expected for the diffuse reflectance as a function of the medium absorption/reduced-scattering property was observed for four conditions of the glucose-medium composition that differed either in the effective glucose concentration or the host medium scattering property. The cumulation of glucose up to 17.8mg/dL and 8.9mg/dL in the host medium having the same optical properties resulted in linear regression slopes of 0.0032 and 0.0030, respectively. The cumulation of the glucose up to 17.8mg/dL in an aqueous host medium that differed two folds in the reduced scattering property caused the linear regression slope to differ between 0.0032 and 0.0019. The $R^2$ values of all cases were all greater than 0.987.

## 1. Introduction

There is a significant and enduring demand for continuous and convenient monitoring of glucose in a substantial proportion of population for a broad spectrum of pathological conditions. National diabetes statistics report for 2020 show that 34.2 million (~10%) American have diabetes, and 88 million American adults (~33%) have prediabetes [1]. With the increase of the obesity rate in the US population, it can be expected that more American adults will develop prediabetes and diabetes. The prevalence of diabetes is no different at global level; approximately 463 million adults are living with diabetes in the world [2]. The American population diagnosed with diabetes — needing monitoring of glucose level on a regular basis — is estimated to be 26.8 million people — or 10.2% of the population. Frequent glucose monitoring is needed for treating diabetic complications such as hyperglycemia or hypoglycemia. Frequent glucose monitoring to control the extent of blood glucose swings is also important to mitigating long-term secondary effects of diabetes, such as retinopathy, nephropathy, neuropathy and microangiopathy/macroangiopathy [3].

The conventional electrochemical technique to measure glucose level requires taking blood samples by using a finger-prick device from subject. The collected blood samples can be either analyzed by labs in clinics or hospitals or by a small glucometer for home use. Although the direct enzymic measurement of serum glucose assures accurate information, the invasive needle-pricking step to collect blood often results in patients (particularly young patients) being reluctant to adopt the process [4] to meet the needs of frequent monitoring, such as after food intake. The discreteness of the invasive sampling approach also makes it difficult to apply to continuous glucose monitoring during sleep and exercise [5-8].

Recently, researchers and industry have shown great interest for developing non-invasive methods/devices for glucose assessment [9]. The non-invasiveness of these methods requires that a measurement probed by an on-surface or off-surface sensor shall respond to the alteration of tissue properties caused by the change of glucose concentration in serum or interstitial spaces [10] in comparison to a calibratable baseline. Non-invasive methods can belong in two categories: non-optical and optical [9]. The three well-known non-optical methods are: (1) electromagnetics sensing (the glucose level is proportional to the ratio between input and output voltages (or currents) of two inductors where the subject's skin is placed) [11], (2) bioimpedance spectroscopy, a.k.a. dielectric impedance spectroscopy (the changes in the permittivity and conductivity of the membrane in red blood cells is caused by glucose changes) [9, 12], and (3) ultrasound (the propagation time of ultrasound waves through the skin is proportionally affected by the concentration of glucose in the tissue) [13]. Optical non-invasive methods take advantages of the glucose-induced change of polarization, reflectance, scattering and absorption of light when passing through a biological medium [9]. Some of the popular non-invasive optical glucose monitoring methods are: photoacoustic NIR and mid-infrared (MIR) spectroscopy [14], Raman spectroscopy [15, 16], optical polarimetry [17], optical coherence tomography [18, 19], fluorescence spectroscopy [20, 21], and far-infrared (FIR) spectroscopy, as are referred to in two recent comprehensive review articles [8, 9]. The subject's targeted skin area, essential to optical probing, depends on the method used. Potential areas include, but are not limited to, fingers, arms, hands, lips, and ear lobes.

Although these methods are non-invasive, they still required body contact applicators, i.e., the applicator for delivering the probing light to tissue and collecting that light has to touch the skin, hence some of them are called wearable non-invasive glucose sensors [6]. Given the link between human's health and glucose level and its variations, there is a need to discreetly and non-intrusively monitor this physiological indicator in human subjects who are susceptible to diabetes and may be vulnerable due to other health or mental conditions, as they carry out their daily routine including sleep. This has prompted the need for sensing methods that can monitor glucose level in a non-contact fashion without impairing the normal activities of a subject. In [22], the authors presented a non-contact method, in which vibrations of laser-illuminated skin (at a 50cm distance) were recorded by a camera, and then processed by using image processing algorithms to estimate glucose level (compared to a reference point) [22]. The technique tracked temporal changes of reflected secondary speckle produced in human skin (wrist) when being illuminated by a laser beam. A temporal change in skin's vibration profile generated due to blood pulsation was projected correlating with the glucose concentration. In [23], an improved method was proposed to estimate the glucose concentration by tracking the secondary speckle pattern of a laser beam illuminating the human skin near blood artery and modulated by a magnetic field. The magnetic field created Faraday rotation of the linearly polarized light when passing through a medium, which was used as an additional modulating factor to the detected speckle field of which the change was projected to correlate with glucose concentration. Although these two approaches were non-contact, such methods may be subject to two major concerns: 1) usage of laser light and the accompanied eye-safety concern, and 2) usage of camera (imaging)-based methods and the accompanied privacy concern.

Whichever light-based method is used, the approach needs to probe a tissue property that can cause a measurable change of the light due to the change of glucose. And the more direct the light interacts with the glucose molecules, the more sensitive and accurate the measurement would become. In this work, we are interested in non-contact assessment of glucose concentration in tissue over a meter-scale projection distance for the prospect of long-range continuous monitoring with autonomous tracking but without privacy concerns. This limits the technical options to only a few, including diffuse reflectance spectroscopy (DRS). Practically, however, assessing the change to light absorption by glucose over a meter-scale distance using the spectral specificity available in DRS introduces several difficulties that could be challenging to address simultaneously. Two such challenges are: (1) the light collected remotely in response to remote illumination shall have propagated within the tissue, i.e., transdermal [24], rather than diffusely reflected from the surface such that the diffusely remitted light carries the information of absorption by glucose deeper in the tissue; (2) the light intensity detected remotely in response to remotely applied eye-safe (thus low spectral intensity) illumination shall be adequate to allow a measurement time-scale that is amenable to practical use. To enhance the signal response to the glucose absorption, it is desirable to collect photons that have propagated farther within tissue, which requires greater sou rce-detector separation (SDS). However, greater SDS and longer photon paths would lower the light remission intensity making detection too costly in time or photon budget. This paper demonstrates DRS in a center-illumination-area-detection geometry that may strike a balance between enhancing sensitivity and reducing collection time using a particular set of projection optics and source capacity, as analyzed in the next section. The rest of the paper is structured as follows. In Section II, the theoretical analysis is provided. Experimental methods are given in Section III. Next, the results are presented in Section IV. Finally, concluding remarks are drawn in Section V.

## 2. Theory
### 2.1. Diffuse reflectance at a meter-scale distance in a center-illumination-area-detection (CIAD) geometry

Consider the general scenario of using light that has diffusely propagated in tissue to assess a tissue constitute (e.g. glucose molecules) based on a contrast of spectral absorption over the background medium, as illustrated in Fig. 1(A). A spatially impulsive illumination of an ideally spectrally flat steady-state (the treatment is limited to steady-state condition in this work) source with a unitary spectral intensity that is applied normal to the medium at a position $\vec{r}'$ is represented by $I_0(\vec{r},\lambda) = \delta(\vec{r}-\vec{r}')\delta(1-\hat{s}\cdot\hat{z})$, where $\hat{z}$ denotes the direction of normal injection to the medium, and $\hat{s}$ is a direction of the light irradiance. The diffusely propagated light remitted at a position $\vec{r}$ at the wavelength $\lambda$ with a mean pathlength in medium that is substantially greater than the reduced-scattering step-size [25, 26] of the medium can be simplified as the following:

$$R(\vec{r},\lambda) = R_{scat}[\mu_s'(\lambda)]exp[-\mu_a(\lambda)\langle l(\lambda)\rangle], \tag{1}$$

where $\mu_a$ [cm$^{-1}$] is the absorption coefficient, $\mu_s'$ [cm$^{-1}$] is the reduced scattering coefficient, $R_{scat}$ is the photon-energy scaled diffuse photon irradiance [cm$^{-2}$·sr$^{-1}$] at the absence of absorption, and $\langle l \rangle$ is the mean pathlength [cm] of the light in tissue between the point-of-incidence (POI) and the point-of-remission (POR). The line-of-sight distance between POI and POR is the source-detector separation (SDS) referred to in the Introduction.

Using diffusely reflected light to assess or monitor the glucose level in tissue concerns assessing the change of $R(\vec{r},\lambda)$ with respect to a baseline calibration. Assuming a stable $R_{scat}$, Eq. (1) changes to

$$\Delta[\mu_a(\lambda)] = \Delta\left[\frac{1}{\langle l(\lambda)\rangle}\cdot ln\left\{\frac{R_{scat}[\mu_s'(\lambda)]}{R(\vec{r},\lambda)}\right\}\right] \tag{2}$$

If the wavelength chosen would render that the light absorption by tissue is dominated by water, and the concentrations of light absorbers in tissue other than glucose molecules remain stable over the course of measurement, the change to the spectral absorption of tissue with respect to the normal value is then attributed to the change of glucose only [27, 28] as

$$\Delta[\mu_a(\lambda)] = \varepsilon_{glu}(\lambda)\cdot\Delta[glu], \tag{3}$$

where $\varepsilon_{glu}(\lambda)$ is the molar extinction coefficient of glucose, and $[glu]$ is the molar concentration of glucose averaged over the interstitial [10] and vascular spaces as a first-order approximation. The application endpoint is to retrieve $\Delta[glu]$, which is dictated by the ability to resolve $\Delta[\mu_a(\lambda)]$, which in this work is assessed by diffuse reflectance of a unique configuration.

Here, the important element of non-contact assessment of glucose concentration in tissue is the meter-scale projection distance. This projection distance, being significantly greater than the SDS that is practical on an exposed area of tissue, essentially limits the projection optics to parallel-to-the-axis. Therefore, what can be collected from the tissue for the assessment is limited to the light flux that is within a very small solid angle centered on a direction parallel to the optical axis of the detection geometry. Due to this geometric confinement, the analysis of the diffuse reflectance from tissue in this long-distance geometry assesses only the flux of the light remission that is normal to the tissue surface in response to a pencil beam illumination injected normal to the medium surface.

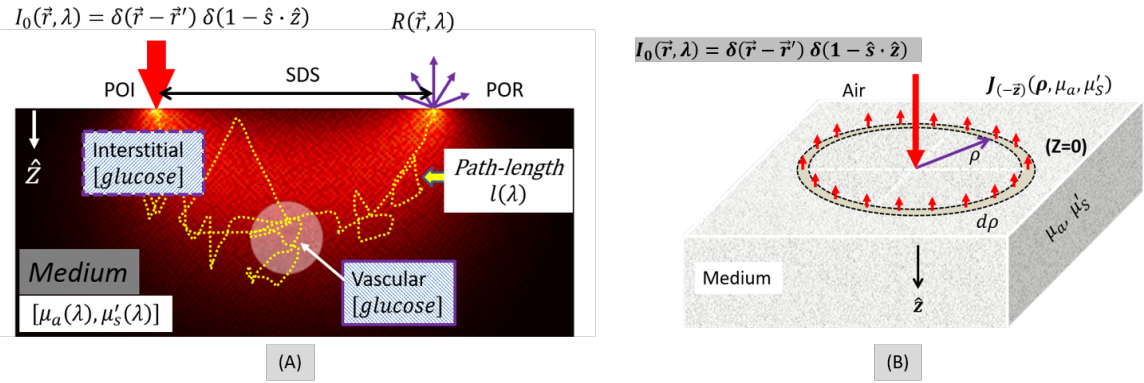

Fig. 1. (A) The use of light diffusely propagated in tissue to assess the glucose level involves identifying the change to the spectral intensity of the remitted light due to glucose in the vascular and interstitial spaces. The assessment of the glucose-induced change to the diffusely remitted light favors longer pathlength of the light in tissue to enhance the response to the same change of the glucose concentration. (B) In the case that light projection and light collection are performed over a long distance, only the flux of the diffusely reflected light that is within a very small spatial angle can be collected.

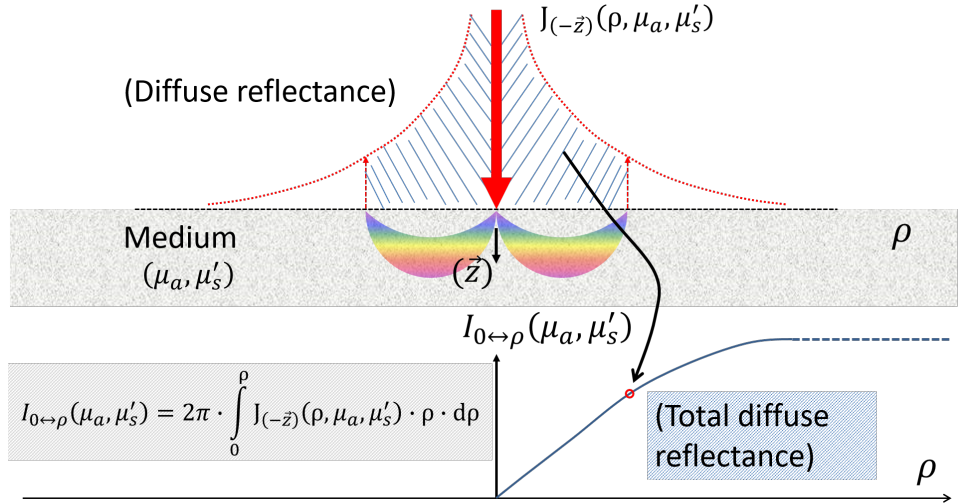

Fig. 2. The upper panel depicts the spatially resolved diffuse reflectance flux $J_{(-\vec{z})}(\rho,\mu_a,\mu_s')$ from a medium of semi-infinite geometry as a function of the SDS $\rho$. As $\rho$ increases, $J_{(-\vec{z})}(\rho,\mu_a,\mu_s')$ decreases rapidly. The lower panel illustrates the total diffuse reflectance $I_{0\leftrightarrow\rho}(\mu_a,\mu_s')$ over the entire circular area with the radius set by $\rho$. The circle on the curve of the lower panel corresponds in numerical value to the shaded portion of the upper panel indicating an integration. As $\rho$ increases, $I_{0\leftrightarrow\rho}(\mu_a,\mu_s')$ will eventually reach a plateau.

### 2.2. Total diffuse reflectance associated with a CIAD geometry projected over a substantial distance

We consider the light diffusely reflected from the surface of a homogeneous turbid medium bounded with air as shown in Fig. 1(B). The POI is set as the origin of the polar coordinate of which the $\vec{z}$ aligns with the direction of light incidence. The spatially resolved diffuse photon flux on the surface at an SDS of $\rho$ from the POI and normal to the tissue surface (i.e., along the $-\vec{z}$ direction) should be azimuthally uniform with respect to the POI for a homogeneous medium and is denoted by $J_{(-\vec{z})}(\rho,\mu_a,\mu_s')$. We assess the entirety of $J_{(-\vec{z})}(\rho,\mu_a,\mu_s')$ over a circular area centered on the POI and having the radius of $\rho$, as the total diffuse reflectance to be collected by the projection optics at a distance that is long enough to render the collection of only the flux aligned with $-\vec{z}$. This total diffuse reflectance for a homogeneous medium is the integration of $J_{(-\vec{z})}(\rho,\mu_a,\mu_s')$ as

$$I_{0\leftrightarrow\rho}(\mu_a,\mu_s') = 2\pi \cdot \int_0^\rho J_{(-\vec{z})}(\rho,\mu_a,\mu_s') \cdot \rho \cdot d\rho. \tag{4}$$

As $\rho$ increases, $J_{(-\vec{z})}(\rho,\mu_a,\mu_s')$ will decrease rapidly due to diffuse attenuation by combined tissue scattering and absorption, meaning the total light remission will not increase much after $\rho$ reaches a threshold (or equivalently the scattering step-size scaled dimensionless term $\mu_s'\rho$). This conceptual pattern of the decrease of $J_{(-\vec{z})}(\rho,\mu_a,\mu_s')$ associated with a pencil beam illumination is illustrated in the upper panel of Fig. 2. Equivalently, as $\rho$ increases, $I_{0\leftrightarrow\rho}(\mu_a,\mu_s')$ would increase until reaching a plateau since the total light of diffuse remission collected on the surface cannot exceed the light injected into the medium.

The total photon flux along the $-\vec{z}$ direction over the CIAD area having a radius of $\rho$ can be decomposed to two parts as:

$$I_{0\leftrightarrow\rho}(\mu_a,\mu_s') = I_{0\leftrightarrow\infty}(\mu_a,\mu_s') - I_{\rho\leftrightarrow\infty}(\mu_a,\mu_s'), \tag{5}$$

where $I_{0\leftrightarrow\infty}(\mu_a,\mu_s')$ corresponds to the total photon flux collected over the entire half-plane of the medium-air interface, and $I_{\rho\leftrightarrow\infty}(\mu_a,\mu_s')$ is the total photon flux collected outside the CIAD area of the radius $\rho$. As $\rho$ or equivalently $\mu_s'\rho$ increases, $I_{\rho\leftrightarrow\infty}(\mu_a,\mu_s')$ shall approach 0, and $I_{0\leftrightarrow\rho}(\mu_a,\mu_s')$ would approach $I_{0\leftrightarrow\infty}(\mu_a,\mu_s')$

(which is bounded at the input light intensity subtracting the amount of absorption by the tissue) and thus may be approximated by $I_{0\leftrightarrow\infty}(\mu_a, \mu_s')$ above a threshold of $\mu_s'\rho$. In the following, we demonstrate how $I_\rho(\mu_a, \mu_s')$ depends upon $\mu_a$ and $\mu_s'$ or $\mu_s'\rho$, while identifying an approximate relation between the medium's absorption-to-reduced scattering ratio and the total diffuse reflectance acquired in the CIAD geometry.

## 2.3. Model of total diffuse reflectance associated with a CIAD geometry projected over a substantial distance

### 2.3.1. Diffuse photon flux along the $-\vec{z}$ direction associated with an infinite homogeneous medium

We first consider the diffuse photon flux in a homogeneous turbid medium of infinite geometry as is illustrated in Fig. 3(A) when illuminated by an isotropic point steady-state source S of unitary power. The isotropic source S is set at $(0,0,z_a)$, where $z_a = (\mu_s')^{-1}$, of which $\mu_s' = \mu_s(1-g)$, $\mu_s$ is the scattering coefficient [cm$^{-1}$] and $g$ is the anisotropy factor of tissue scattering. We further consider the tissue dimension for the diffuse reflectance to be probed remotely by an optical projection setup that can be configured with the popular 1" cage system. A parallel-to-the-axis condition pertinent to a 1" cage optics would result in an effective tissue dimension of approximately 2 cm in diameter at parallel-to-the-axis projection. The photon fluence rate at a position $(\rho, 0, 0)$ which has a distance of

$$l_{real} = \sqrt{\rho^2 + z_a^2}, \tag{6}$$

with respect to the source $S(0,0,z_a)$ is [29]

$$\Psi(\rho, 0, 0) = \frac{1}{4\pi D} \frac{1}{l_{real}} exp(-\mu_{eff} l_{real}), \tag{7}$$

where $D = 1/[3(\mu_a + \mu_s')]$ is diffusion coefficient [cm] and $\mu_{eff} = \sqrt{\mu_a/D}$ is effective attenuation coefficient [cm$^{-1}$]. The corresponding photon flux at the position $(\rho, 0, 0)$ and projected along the $-\vec{z}$ direction is [29]

$$J_{(-\vec{z})}(\rho, 0, 0, \mu_a, \mu_s') = -D(-\vec{z}) \cdot \nabla\Psi(\rho, 0, 0) = \frac{1}{4\pi} \frac{z_a(\mu_{eff} l_{real} + 1)}{(l_{real})^3} exp(-\mu_{eff} l_{real}). \tag{8}$$

### 2.3.2. Diffuse photon flux associated with a homogeneous medium of semi-infinite geometry

We consider the diffuse photon flux in a homogeneous turbid medium of semi-infinite geometry as is illustrated in Fig. 3(B) in response to a directional point illumination injected normal to the medium surface. The geometry of (B) shares the same polar coordinates and dimensions with (A) when pertinent. The directional point source normal to the medium surface is applied at the origin of the coordinate. The diffuse photon flux caused by the directional source is to be equivocated by an isotropic source positioned at $(0,0,z_a)$ [29]. The photon flux along the $-\vec{z}$ direction is assessed at the position $(\rho, 0, 0)$ that has a distance of $l_{real}$ defined in Eq. (6) from the equivalent isotropic source at $(0,0,z_a)$ that is placed inside the medium. The effect of the medium-air boundary is treated with a Robin-type extrapolated-zero boundary condition [30], by setting an image of the equivalent isotropic source with respect to an imaginary boundary that is set at a distance of $z_b$ away from the physical boundary. The distance of the extrapolated boundary from the physical tissue-air boundary is $z_b = 2AD$ [cm], where $A = (1+\xi)/(1-\xi)$ and $\xi = -1.44 n_{tiss}^{-2} + 0.710 n_{tiss}^{-1} + 0.668 + 0.0636 n_{tiss}$, where $n_{tiss}$ is the refractive index of the medium that is set at 1.4 in this work.

The image source at $[0,0,-(z_a + 2z_b)]$ incurs another distance for the point $(\rho, 0, 0)$ on the medium-air boundary as:

$$l_{imag} = \sqrt{\rho^2 + (z_a + 2z_b)^2} \tag{9}$$

The photon fluence rate and photon flux assessed on $(\rho, 0, 0)$ now will contain the contribution by both the equivalent isotropic source at $(0,0,z_a)$ and the image of it at $[0,0,-(z_a + 2z_b)]$ with respect to the extrapolated boundary. The two components of photon fluence rate on $(\rho, 0, 0)$ contributed by the two sources have opposite signs, but the two components of photon flux on $(\rho, 0, 0)$ and along the $-\vec{z}$ direction contributed by the two sources have the same sign. The composite photon flux along the $-\vec{z}$ direction when assessed at $(\rho, 0,0)$ is thus

$$J_{(-\vec{z})}(\rho,0,0,\mu_a,\mu_s') = \frac{1}{4\pi}\left[\frac{z_a(\mu_{eff}l_{real}+1)}{(l_{real})^3}exp(-\mu_{eff}l_{real}) + \frac{(z_a+2z_b)(\mu_{eff}l_{imag}+1)}{(l_{imag})^3}exp(-\mu_{eff}l_{imag})\right] \quad (10)$$

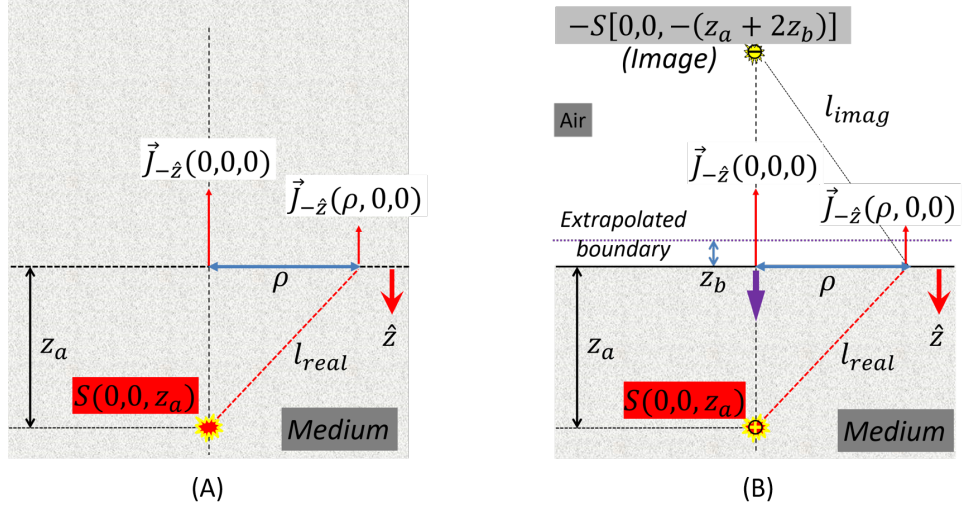

Fig. 3. (A) A turbid medium of infinite geometry. An isotropic point source is set at $(0, 0, z_a)$. The photon flux along the $-\vec{z}$ direction is assessed at a position $(\rho, 0, 0)$ which has an SDS of $l_{real}$ from the source at $(0, 0, z_a)$. (B) A turbid medium of semi-infinite geometry. A directional point source normal to the medium surface is applied at the origin of the polar coordinate. The diffuse photon flux caused by the directional source is to be equivocated by an isotropic source positioned at $(0, 0, z_a)$ as in (A). The photon flux along the $-\vec{z}$ direction assessed at a position $(\rho, 0, 0)$ on the medium surface that has a SDS of $l_{real}$ from the equivalent isotropic source at $(0, 0, z_a)$ will contain the effect of the medium-air boundary which is treated by using an image source approach. The image source is the mirror image of the equivalent isotropic source with respect to an extrapolated boundary set at a distance of $z_b$ away from the physical boundary.

### 2.3.3. Total photon flux from a homogeneous medium of semi-infinite geometry in the CIAD geometry

Substituting Eq. (10) to Eq. (4) gives the total photon flux along the $-\vec{z}$ direction over the CIAD area with a radius of $\rho$. For the convenience of the subsequent analyses, the total photon flux $I_{0\leftrightarrow\rho}(\mu_a,\mu_s')$ is split to two parts by referring to Eq. (5) as:

$$I_{0\leftrightarrow\infty}(\mu_a,\mu_s') = 2\pi\int_0^\infty J_{(-\vec{z})}(\rho,0,0,\mu_a,\mu_s')\cdot\rho\cdot d\rho$$
$$= \frac{1}{2}\{\exp(-\mu_{eff}z_a) + \exp[-\mu_{eff}(z_a+2z_b)]\}, \quad (11)$$

$$I_{\rho\leftrightarrow\infty}(\mu_a,\mu_s') = 2\pi\int_\rho^\infty J_{(-\vec{z})}(\rho,0,0,\mu_a,\mu_s')\cdot\rho\cdot d\rho$$
$$= \frac{1}{2}\left\{\frac{z_a}{\sqrt{\rho^2+(z_a)^2}}\exp\left(-\mu_{eff}\sqrt{\rho^2+(z_a)^2}\right) + \frac{z_a+2z_b}{\sqrt{\rho^2+(z_a+2z_b)^2}}\exp\left(-\mu_{eff}\sqrt{\rho^2+(z_a+2z_b)^2}\right)\right\}. \quad (12)$$

Equation (11) evolves to

$$2I_{0\leftrightarrow\infty}(\mu_a,\mu_s') = exp\left[-\frac{\mu_{eff}}{\mu_s'}\right] + exp\left[-\frac{\mu_{eff}}{\mu_s'}\left(1+\frac{2z_b}{z_a}\right)\right]. \quad (13)$$

Where

$$\frac{2z_b}{z_a} = \frac{4A}{3}\frac{1}{\left(1+\frac{\mu_a}{\mu_s'}\right)}. \quad (14)$$

And Eq. (12) can be rewritten, by using a dimensionless term $\mu_s'\rho$ which specifies the relative size of the CIAD area with respect to the reduced scattering step-size, as the following:

$$2I_{\rho\leftrightarrow\infty}(\mu_a,\mu_s') = \frac{1}{\sqrt{(\mu_s'\rho)^2+1}}\exp\left[-\frac{\mu_{eff}}{\mu_s'}\sqrt{(\mu_s'\rho)^2+1}\right] + \frac{1+\frac{2z_b}{z_a}}{\sqrt{(\mu_s'\rho)^2+\left(1+\frac{2z_b}{z_a}\right)^2}}\exp\left[-\frac{\mu_{eff}}{\mu_s'}\sqrt{(\mu_s'\rho)^2+\left(1+\frac{2z_b}{z_a}\right)^2}\right].$$

(15)

It is apparent from Eq. (15) that as $\mu_s'\rho \to \infty$, $I_{\rho\leftrightarrow\infty}(\mu_a,\mu_s') \to 0$, which will cause $I_{0\leftrightarrow\rho}(\mu_a,\mu_s')$ to reach $I_{0\leftrightarrow\infty}(\mu_a,\mu_s')$. Using the following notation

$$\eta = \frac{\mu_a}{\mu_s'},$$

(16)

it is straightforward to demonstrate that

$$I_{0\leftrightarrow\infty}(\eta) = 1 - \sqrt{3\eta(\eta+1)}\left(1+\frac{2A}{3}\right),$$

(17)

and by using the following conversion

$$I_{conv} = \frac{1}{1+\frac{2A}{3}}[1 - I_{0\leftrightarrow\infty}(\eta)],$$

(18)

we have

$$\eta = \frac{\mu_a}{\mu_s'} = \frac{1}{2}\left[\sqrt{1+\frac{4}{3}(I_{conv})^2} - 1\right].$$

(19)

Experimental arrival of the simple relation of Eq. (19) or alike between the medium's absorption/reduced scattering ratio and the modified diffuse reflectance would render in principle the glucose concentration change be resolved with the identification of the scattering properties of the medium. We hereby also note that the algebraic simple equation (17) relating the total diffuse reflectance with the tissue absorption/reduced scattering properties differs from the well-known equation of Kubelka-Munk (KM) model [31] of the following :

$$\eta_{KM} = \frac{2\mu_a}{\frac{3}{4}\mu_s(1-g)-\frac{1}{4}\mu_a(1-3g)} = \frac{[1-I_{KM}(\mu_a,\mu_s,g)]^2}{2I_{KM}(\mu_a,\mu_s,g)}$$

(20)

Where $I_{KM}(\mu_a,\mu_s,g)$ is the collimated reflected light under a diffuse illumination. Equation (20) leads to

$$I_{KM}(\mu_a,\mu_s,g) = (1+\eta_{KM}) - \sqrt{(1+\eta_{KM})^2 - 1}$$

(21)

### 3. Experimental Methods

This section details (1) the experimental configurations associated with the non-contact diffuse reflectance system that operated in the CIAD geometry, (2) the sample preparations for validating the theoretical model connecting the tissue properties with the total diffuse reflectance, (3) the identification of a spectral window of approximately 200 nm that may be optimal for CIAD-based assessing of glucose in an aqueous turbid medium based on absorption changes to the diffuse reflectance, and (4) the demonstration of resolving glucose change of up to 17.83mg/dL concentration in an aqueous turbid medium at the distance of ~1.1 meters. In terms of examining the aforementioned theoretical model that relates the tissue properties with the total diffuse reflectance, the total diffuse reflectance was assessed at varied dimension of the collection area from a solid medium rendering ideally fixed sample properties, varying the scattering of an aqueous turbid medium when probed at a constant dimension of the collection area, and varying the absorption of an aqueous turbid medium when probed at a constant dimension of the collection area. In terms of examining the potential of the total diffuse reflectance in assessing the change of glucose concentration in an aqueous turbid medium, the measurement was performed for four sets of conditions all rendering the glucose concentration to vary over three orders of magnitude by using control solutions of different effective

## 3.1. Non-contact diffuse reflectance system with a meter-scale working distance configured in CIAD geometry

The non-contact diffuse reflectance system with a working distance of approximately 1.1 meters is schematically illustrated in Fig. 4. A high-brightness broadband laser-driven light source (LDLS) (EQ-99XFC, Energetiq Inc., Woburn, MA) [32-34] was used for illumination. This LDLS light source had a spectral power greater than 10µW/nm over 190–980nm, and a spectral power lower than 10µW/nm when extending to 2100nm, according to the vendor datasheet. The ultrabroad bandwidth of this light source also facilitated in-house transmission-geometry spectrophotometry (see Section 3.2) for the identification of a spectral window of approximately 200nm around $1.2 \mu m$ for assessing the light absorption due to glucose in water. The LDLS was coupled via a 200µm 0.22 NA low-OH fiber (M25L02, Thorlabs, Newton, NJ) to a 1" cage system for collimated light delivery. The collimated projection was configured by using an achromatic lens (sL1, focal length=8.0mm, C240TME-B, Thorlabs, Newton, NJ) and an achromatic relay lens (sL2, focal length=250mm, LA1461-C, Thorlabs, Newton, NJ). A linear polarizer (sP) mounted on a rotational stage was added after sL2. The optical path from the fiber-launching adaptor to the linear polarizer was contained within 1" lens tubes. The collimated beam was projected horizontally to a 2"×2" mirror (M) located at approximately 1.0 meters distance to illuminate at slightly off the normal angle on the sample placed approximately 0.1 meters below the mirror. The total distance from the illumination-path polarizer to the sample was approximately 1.1 meters. The light of diffuse reflectance from the sample was deflected by the same mirror M to another 1" cage system configured also at approximately 1.0 meters distance from the mirror, for coupling to another fiber patch-cable of 200µm and 0.22NA. The total distance from the sample to the detection-path polarizer was also approximately 1.1 meters. The return path of the light for detection contained a linear polarizer (dP) to mitigate the specular reflectance from the medium surface, an achromatic relay lens (dL2, identical to sL2), and an achromatic focusing lens (dL1, focal length=15.29mm, C260TME-B, Thorlabs, Newton, NJ). The optical path from the linear polarizer to the fiber-launching adaptor was also contained within 1" lens tubes. The detection fiber was adapted to the entrance slit of a monochromator (Princeton Instruments, sp300i) that was mounted with three gratings, of which the one with the lowest groove density of 150-grooves/mm and blazed at 600nm was used for this work [35]. The monochromator had an attached intensified 16-bit CCD camera (PI-MAX II, Acton Research) [36] cooled to -20°C with a chip of 512×512 elements totaling a sensing area of $12.0 \times 12.0\ mm^2$. All 512 vertical pixels at one horizontal pixel were pinned to perform at the spectroscopy mode with the maximum pixel read-out. Approximately 200nm spectral coverage was rendered. The control of the CCD exposure time and gain as well as the monochromator center wavelength were performed by using WinSpec interface [36]. The data acquired was post-processed in Matlab environment. The data acquisition from turbid medium (solid or aqueous based) was performed consistently at 10 seconds of the CCD exposure time at the same gain setting set to accommodate monitoring the changes of the diffuse reflectance signal from samples with the optical properties varying over a great range. The effect of ambient lighting was minimized by shielding the entire optical paths and dimming the room light.

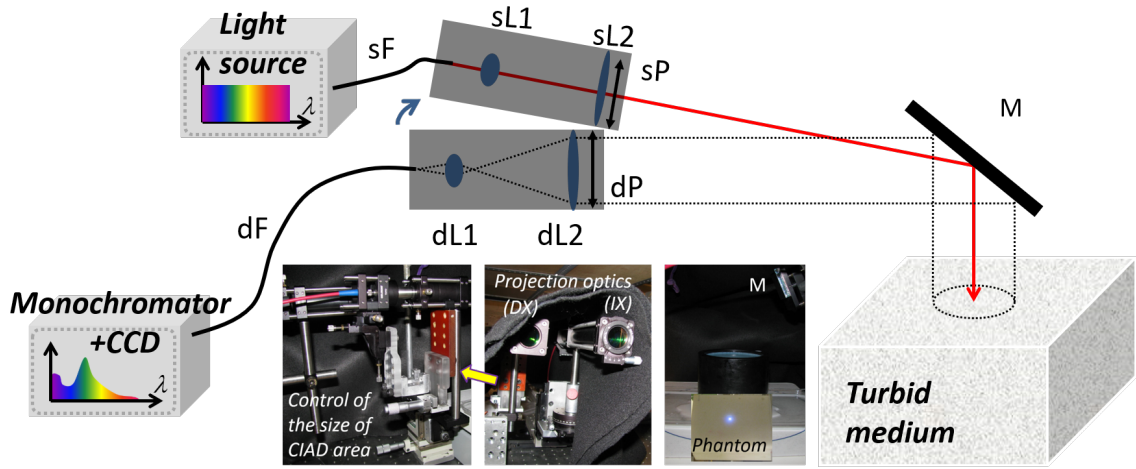

Fig. 4. Upper panel: DRS system. Components used in the source channel: sF, source or illumination fiber; sL1, lens 1; sL2, lens 2; sP, linear polarizer. Components used in the detection channel: dF, detection fiber; dL1, lens 1; dL2, lens 2; dP, linear polarizer. M, mirror. IX: Illumination. DX: detection. Lower panel: left, the setup used to control the size of the detection area with repeatability; middle, the illumination path (IX) and detection path (DX) enclosed within lens tubes; right, the mirror for deflecting the light coming from the left to the subject of assessment located below the mirror and placed on a stirring machine. Two kinds of subjects are shown, one is aqueous sample housed by a container with a 2.75" magnetic stirrer at the bottom for agitation, another one is solid phantom placed directly on the stirring machine used as a platform. The illumination light spot on the solid tissue phantom is visible.

### 3.2 Transmission spectrophotometry to identify a spectral window for diffuse reflectance assessment of glucose in aqueous medium

Fig. 5 shows the schematic of a 1"-cage transmission setup developed in-house for assessing the spectral absorption difference between glucose and water over 1.1 µm to 2.1 µm. The broad-band light from the LDLS source described in the previous section coupled by a 200 µm fiber was collimated by an aspherical lens (focal length=8.0 mm, C240TME-B) and projected onto a 50 µm fiber via focusing by another aspherical lens (focal length=15.29 mm, C260TME-B). The beam passing a cuvette of 1cmX1cm cross-section was aligned by two pinholes set at the far opposite sides of the cuvette. A custom platform with cuvette holding mechanism was fabricated to maintain a fixed position of the cuvette after interchanged for different samples. The pinholes were also fixed in position to keep the beam collimation as consistent as possible. Lens tubes were used to shield the light path when feasible, and the entire optical path was covered with black cloths. The light transmitted through the water or glucose sample housed in the cuvette of 1 cm optical path and collected by the 50 µm fiber was coupled to the entrance slit of the monochromator of Fig. 4. For each sample of the cuvette placed in the transmission beam path, the center wavelength of the monochromator was changed from 1200 nm to 2000 nm at a step of 200 nm, to cover a total spectral range of approximately 1.1 µm to 2.1 µm. The CCD exposure time and gain settings were fixed while allowing acquisition of the collimated light over the entire spectral coverage from 1.1 µm to 2.1 µm without saturation. One cuvette contained water and another cuvette contained glucose at a concentration of 25 g/dL (ReliOn Glucose Shot, Walmart Pharmacy). The spectral scanning of the transmission beam over the 1.1 µm to 2.1 µm range at a step of 200 nm was completed on each sample, before switching the sample within the fixture and resetting the same beginning (at 1.1 µm) spectral range of the monochromator to keep the same direction of scanning the center-wavelength. The objective was to identify a spectral window over the 2$^{nd}$ near-infrared window that presented the following spectral characteristics considered desirable for sensing glucose in tissue medium abundant of water: glucose is more absorbing than water while water has relatively low absorption. These two features would allow longer pathlength of light in tissue to make the diffuse reflectance more sensitive to the change by glucose absorption. Since the measurement was to identify the relative spectral absorption between water and glucose, the glucose sample was tested at only one concentration.

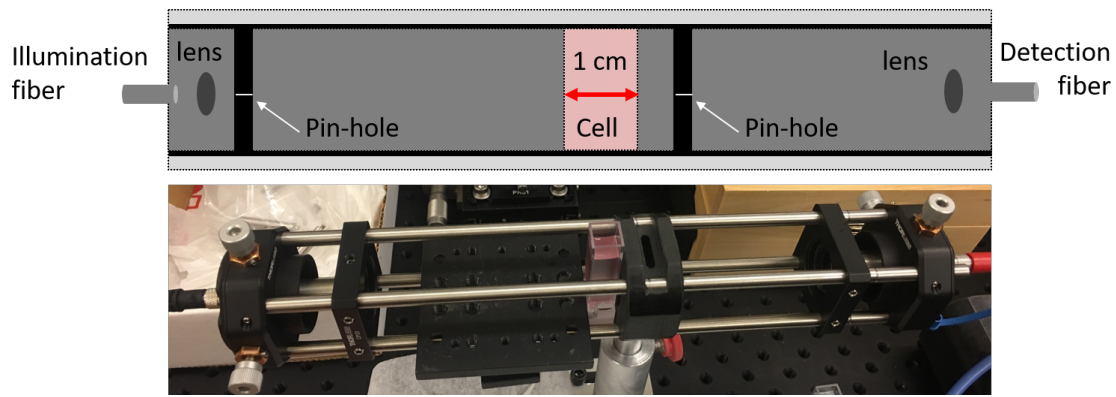

Fig. 5. Upper panel: Schematic diagram of the 1"-cage transmission setup for assessing the spectral absorption difference between glucose and water over 1.1um to 2.1um. The broad-band light illuminated by a 100um fiber was collimated by an aspherical lens of focal length 8.0mm and projected a detection fiber via another aspherical lens of focal length 15.29mm. The beam passing a cuvette of 1cmX1cm cross-section was aligned by two pinholes set at the opposite sides of the cuvette. A custom platform was fabricated to maintain a fixed position of the cuvette when interchanged for different samples, and the tight contact of the pinholes with the cuvette to sustain the consistency of the beam after switching the samples. Lower panel: The optical setup of the in-house transmission spectrophotometry with the lens tubing removed and covering cloth removed.

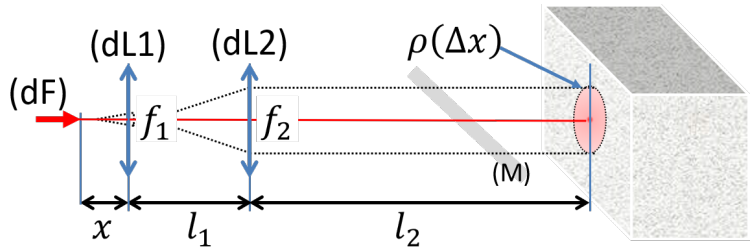

Fig. 6. Schematic for the ray tracing to estimate the change of the collection area as the distance between the collection fiber dF and tis focusing lens dL1 was adjusted. The distance between the detection fiber dF and the focusing lens dL1 of a focusing depth of $f_1$ is $x$, the distance between the focusing lens dL1 and the relay lens dL2 of a focusing depth $f_2$ is $l_1$, and the distance between the relay lens dL2 and the sample is $l_2$. The radius of the collection beam on the sample is $\rho(\Delta x)$, where $\Delta x$ is the displacement of the detection fiber dF from the position projecting the smallest spot on the sample. The inset shows the three solid phantoms used for assessing the change of the total diffuse reflectance as a function of the size of the collection of the CIAD geometry.

### 3.3 Approaches to validate the diffuse reflectance in the CIAD geometry

According to the theoretical analysis in Section 2, the diffuse reflectance at a specific wavelength pertinent to the CIAD geometry reveals the following patterns: (1) The total diffuse reflectance acquired from a medium of fixed optical properties will increase as the size of collection increases until reaching a plateau. (2) The total diffuse reflectance configured at a fixed area of collection will increase (until reaching a plateau) as the medium's scattering increases. (3) The total diffuse reflectance configured at a fixed area of collection will decrease as the medium's absorption increases. The following measurements were performed to validate these model-indicated patterns of the diffuse reflectance.

#### 3.3.1. Total diffuse reflectance from a medium of fixed properties as a function of the size of collection

Solid tissue phantoms were used to assess the total diffuse reflectance as a function of the size of collection area. The use of solid phantom helped maintain that the medium properties were identical as the area for remote collection of diffuse reflectance was changed in size while the illumination spot remained at a minimal size of less than 1.0 mm in diameter.

The model-fitting of the total diffuse reflectance requires knowing the size of the collection area. The size of the collection area can be appreciated by reversing the optical beam path in the detection channel; however, it would be impractical to precisely measure the collection beam size on the sample. In order to examine the model prediction of the diffuse reflectance as the size of collection area increased, the change of the diffuse reflectance was assessed by accurately controlling the change of the distance between the collection fiber and the focusing lens.

The configuration to precisely change the distance between the collection fiber and the focusing lens to cause a change of the size of collection area on the sample is illustrated schematically in Fig. 6. The beam path of the detection channel from the mirror to the sample that deflected with respect to the beam path from the collection fiber to the mirror is unfolded to facilitate ray tracing analysis. Under the practical situation, the only way to precisely (and allowing accurate repetition for measuring multiple samples) change the size of the beam area on the sample without perturbing up the optical alignment is to adjust the relative position of the detection fiber with respect to the focusing lens dL1. To facilitate accurate large-scale change of the relative position of the detection fiber with the focusing lens dL1 (as needed to cause a large change of the size of diffuse collection for examining the model prediction), a unique design was implemented for the mechanical supporting portion of the optical path containing the fiber adaptor and the focusing lens. That design of the mechanical supporting can be seen in the left-most photograph of the inset of Fig. 4. The mechanical support engaging 1 miniature 3-way translation stage and a small 1-way translation stage combined in a custom fixation structure allowed the position of the detection fiber with respect to the focusing lens be adjusted over an 8 mm distance with micrometer resolution, without appreciably degrading the optical alignment of the cage system. The detection fiber was first set at a position corresponding to the minimal collection spot size on the sample, then translated at a step of 100 $\mu m$ over an 8 mm distance away from the focusing lens dL1. This distance of translating the fiber corresponded to the collection spot size increasing from the minimal size of approximately 0.5 mm in diameter to approximately 2.0 cm in diameter rendered by the 1" cage system projected at the 1.10 meters distance. As the fiber was displaced from the position projecting the smallest spot size on the sample, the total diffuse reflectance was observed to increase initially till leveling off briefly, followed by a reduction indicating that the light within the collecting aperture of the fiber was cut-off by the optical apertures of the cage system.

Three solid phantoms of grossly rectangular shape (PB0383, PB0385, and PB0335, in-house shaped from the raw materials provided by Biomimic, INO Inc, Québec, Canada) were used [25]. Of these solid phantoms, the absorption was factory calibrated over 650-850nm and the reduced scattering was vendor calibrated at 800nm by using time-resolved measurements. The PB0383 and PB0385 were identical in size, whereas PB0335 was slightly smaller (shorter in the smallest dimension of the rectangular block). The three solid tissue phantoms had nearly identical $\mu_s'$ (10.4cm$^{-1}$ for PB0383, 10.6cm$^{-1}$ for PB0385, and 10.8cm$^{-1}$ for PB0335) but $\mu_a$ that scaled approximately 1:2:4 (0.055cm$^{-1}$ for PB0383, 0.109cm$^{-1}$ for PB0385, and 0.21cm$^{-1}$ for PB0335). Each phantom was placed to have its largest flat surface facing the beam. Each of the two phantoms of the same size (PB0383 and PB0385) was placed directly on a magnetic agitating device (see the following section) used as a platform. The smallest (PB0335) of the three phantoms was placed on the magnetic agitator with a substrate plate to maintain approximately the same height of the phantom surface. The surface of the solid phantom facing the beam was also maintained at approximately the same as the beam-receiving spot on aqueous phantom under magnetic agitation to be described in the following section.

The change of the total diffuse reflectance as a function of the change of the relative position between the collection fiber (dF) and its focusing lens (dL1) was modeled by the following steps, by reversing the beam-path from the collection area on the sample to the collection fiber, as shown in Fig. 6: (1) Assume that a minimal spot size for the light collection on the sample corresponded to a dF-dL1 distance of $x_0$. Then a varying displacement of $\Delta x$ of the collection fiber dF away from the focusing lens corresponded to a varying distance of $x(\Delta x) = x_0 + \Delta x$ between dF and dL1. (2) This varying distance $x(\Delta x)$ was implemented in ray-tracing matrix-analysis to result in a varying radius, $\rho(\Delta x)$, of the collection area on the sample. (3) The $\rho(\Delta x)$ was then plugged to Eq. (15) and combined with Eqs. (13) and (5) to deduce the (relative) change of the total diffuse reflectance, for comparing with experimental measurements.

Denoting the ray direction when leaving the fiber as $r'(x)$, and the off-axis displacement of the ray when leaving the fiber as $r(x)$, the following ray-tracing transmission can be expected:

$$\begin{bmatrix} \rho(x) \\ \rho'(x) \end{bmatrix} = \begin{bmatrix} 1 & l_2 \\ 0 & 1 \end{bmatrix} \begin{bmatrix} 1 & 0 \\ -1/f_2 & 1 \end{bmatrix} \begin{bmatrix} 1 & l_1 \\ 0 & 1 \end{bmatrix} \begin{bmatrix} 1 & 0 \\ -1/f_1 & 1 \end{bmatrix} \begin{bmatrix} 1 & x \\ 0 & 1 \end{bmatrix} \begin{bmatrix} r(x) \\ r'(x) \end{bmatrix}, \tag{22}$$

where $\rho'(x)$ is the ray direction when impinging on the sample surface. Equation (22) evolves to the following:

$$\rho(x) = \left[1 - \frac{l_1+l_2}{f_1} - \frac{l_2}{f_2} + \frac{l_1 l_2}{f_1 f_2}\right] r(x) + \left[\left(1 - \frac{l_1+l_2}{f_1} + \frac{l_2 l_1}{f_2 f_1}\right)x - \frac{l_2}{f_2}(l_1 + x) + l_1 + l_2\right] r'(x). \tag{23}$$

Denoting the radius of the focusing lens dL1 as b, we have the ray direction when exiting the fiber at a lateral position of 100 µm or half of the diameter of the detection fiber as

$$r'(x) = \frac{b-0.1}{x} = \frac{b-0.1}{x_0+\Delta x}, \tag{24}$$

where all displacement entities take a dimension of mm. The estimation of the total diffuse reflectance by Eq. (5) with the implementation of Eq. (23) is based on the postulation that all rays within the numerical apertures of the projection optics could be collected equally well, which was not true. To compensate for the difference in terms of the light collection at different size of the collection area, the Eq. (5) was implemented at a slightly different form as

$$I_{0 \leftrightarrow \rho}(\mu_a, \mu_s') = I_{0 \leftrightarrow \infty}(\mu_a, \mu_s') - I_{\rho \leftrightarrow \infty}(\mu_a, \mu_s') \cdot exp\left[-\left(\frac{\rho(x)}{\rho_0}\right)^2\right], \tag{25}$$

where $\rho_0$ is a reference length used to adjust the contribution of the term to be subtracted that corresponds to the total photon flux collected outside the CIAD area, and the Gaussian form affiliated with $I_{\rho \leftrightarrow \infty}(\mu_a, \mu_s')$ was used to approximate the effect of the non-uniform collection of the beam over the numerical aperture permitted by the collection optics. This modification to the collection beam size was compared with computer simulation based on Zemax as will be presented in the results section.

### 3.3.2. Total diffuse reflectance at ~1.1 meters distance from an aqueous medium as a function of the medium scattering at a fixed size of the detection area

Total diffuse reflectance was acquired from an aqueous sample with the reduced scattering coefficient varying over 2 orders of magnitude. A container of approximately 350ml capacity was fabricated from a black polycarbonate material of 4.125" in the outside diameter. The container had a bottom layer thickness of approximately 0.04" (1mm) to allow a 2.75" magnetic stirrer to be agitated by using an agitating machine (Model 760, VWR High Volume Magnetic Stirrer, Radnor, PA). The agitation was set at 180 rotation-per-minute or 3 rotation-per-second for all experiments performed on aqueous samples. The agitation started with the host medium occupying about 60% of the total capacity of the in-house container to avoid spilling of the liquid at the starting of the agitation, and once the agitation stabilized more host medium was added to the near-full capacity of the container to maintain a consistent curved surface profile of the medium under agitation, as illustrated in Fig. 7.

Total diffuse reflectance was acquired from the medium under agitation at a collection-size corresponding to the maximum total diffuse reflectance determined by the testing in 3.3.1. The reduced scattering coefficient of the aqueous medium was controlled by dissolving bulk intralipid solution of 20% concentration (Baxter, Deerfield, IL) in water as the host medium, by using a 1ml insulation syringe. The adding of the 20% bulk intralipid into the host medium in the container were controlled to reach the following cumulative volume of the 20% intralipid in the aqueous medium: 0.0 to 1 ml at a step of 0.1 ml, 1 ml to 20 ml at a step of 0.5 ml, and 20 ml to 30 ml at a step of 1 ml. These resulted in intralipid concentrations that varied 300 times. The concentration of intralipid was converted to reduce scattering coefficient at ~1.2 μm [37, 38].

### 3.3.3. Total diffuse reflectance at ~1.1 meters distance from an aqueous medium as a function of the medium absorption at a fixed size of the detection area

Total diffuse reflectance was acquired from an aqueous sample with the absorption coefficient varying over 1 order of magnitude, by using the same container/agitation/CIAD setup described in 3.3.2. The host medium was the medium experimented in 3.3.2 that contained 30 ml of 20% intralipid dissolved in water of approximately 320 ml in the initial volume. The absorption coefficient of the aqueous medium was controlled by dissolving Indian ink using a 1 ml insulin syringe. Because the ink did not dissolve well in the solution containing intralipid and tended to float on the surface when dropped on to the surface, after the injection, the liquid was first manually agitated by moving a 3/16" hex drive wrench counterclockwise to go against the clockwise flow of the liquid maintained by the magnetic agitation, then left for stabilization by the magnetic agitation until homogenization of the ink was reached, before data acquisition was commenced.

The ink was added to reach the following cumulative volume in the host intralipid medium: 0.0 to 1 ml at a step of 0.1 ml. Adding more than 1 ml ink was not attempted because the signal didn't vary appreciably as the cumulative volume of ink was greater than 0.7 ml. The concentration of ink was converted to absorption coefficient at ~1.2 μm according to [39].

### 3.4 Assessing glucose dissolved in aqueous tissue medium by using total diffuse reflectance at ~1.1 meters distance

Total diffuse reflectance was acquired from an aqueous turbid medium with the effective glucose concentration varying over three orders of magnitude, using the same container/agitation/CIAD setup described in 3.3.2. The host medium was set at two values of the reduced scattering coefficients while added with glucose control solution of two different concentrations. This resulted in a total of 4 sets of measurement conditions concerning glucose concentration and medium properties to assess the response to the glucose concentration change at the same medium scattering properties and the response to the medium scattering change at the same glucose concentration.

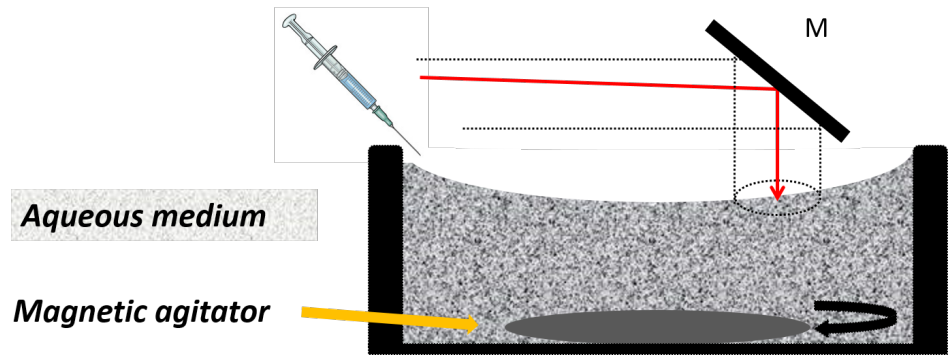

Fig. 7. Configuration of the aqueous sample allowing homogenization of the medium as the medium's scattering or absorption property was changed by injecting scattering or absorption materials by using syringe. A container of approximately 350ml capacity was fabricated from black polycarbonate material of 4.125" in outside diameter. The container had a bottom layer thickness of approximately 0.04" (1mm) to allow a 2.75" magnetic agitator to be activated. The container was concentric to the axis of the rotation of the magnetic agitator. The agitation caused the surface of the aqueous medium to be parabolic, and the light collection area that was centered on the light illumination spot was set at the middle-section between the agitation lowest center point of the aqueous surface and the container edge.

The commercial glucose control solutions used were in 2ml package with an effective glucose concentration of 0.2% in the model 1 and 0.4% in model 2 (AimStrip, Germaine Laboratories, Inc., San Antonio, TX). Five 2 ml bottles of each model were emptied to a small container for administration by using a 1 ml insulin syringe. The four sets of measurements correspond to two values of the reduced scattering coefficient of the host medium, and two values of the glucose concentration in the control solution of the same set of volumes when added to the host medium. The two control solutions were grossly indistinguishable based on color. The control solution was added to reach the following cumulative volume of the control solution of the respective model in the host intralipid medium: 0.00 to 0.1 ml at a step of 0.01 ml, 0.1 to 1 ml at a step of 0.1 ml, and 1 to 10 ml at a step of 0.5ml. Administration of the glucose at a 0.1 ml step was convenient by using the decimal marks of the 1 ml insulin syringe. Administration of glucose at a 0.01 ml step was practiced by forming a droplet at the tip of the syringe. Approximately 10 droplets were managed with a 0.1ml volume that resulted in an estimated 0.01ml volume of each droplet. These resulted in the glucose concentration in the turbid medium host to vary 1000 times. The effective concentration of the glucose in the intralipid was up to 8.91 mg/dL when administered with the control solution of model 1, and up to 17.83 mg/dL when administered with the control solution of model 2, by considering the molecular weight of glucose to be 1.56g per ml.

## 4. Results
### 4.1 Transmission spectrophotometry to identify a spectral window of absorption difference between glucose and water for diffuse reflectance assessment in the CIAD geometry

Fig. 8(A) shows the raw transmission spectra of the water (solid black line) and the glucose shot solution (dashed red line) over a spectral range of approximately 1.1-2.1 µm. Since the monochromator-CCD could cover only approximately 200 nm in each setting of the center wavelength, the spectral profiles spanning the 1.1-2.1 µm range were concatenated by stitching five individual profiles of approximately 200 nm in range with the center wavelength set at respectively 1200 nm, 1400 nm, 1600 nm, 1800 nm, and 2000 nm. As the center-wavelength of the monochromator was changed, so did the autonomous responsivity compensation of the device (monochromator-CCD combined), and that caused the intensities of the same wavelength but acquired at two different center wavelength settings to differ. The stitching of the five profiles of approximately 200 nm in range was done by using the profile corresponding to a center wavelength of 1200 nm as was, then aligning the profile corresponding to a center wavelength of 1400 nm with the former one at 1300 nm followed by repeating similar procedure for the subsequent individual profiles that overlapped at 1500 nm, 1700 nm and 1900 nm.

If we denote the spectral transmission through water as

$$T_{water}(\lambda) = I_{source}(\lambda) exp[-\mu_a^{water}(\lambda)L], \quad (26)$$

where $\mu_a^{water}(\lambda)$ is the spectral absorption of water, and $L$ =1cm is the pathlength of the cell, and denote the spectral transmission through glucose shot solution as

$$T_{glucose}(\lambda) = I_{source}(\lambda) exp[-\mu_a^{glucose}(\lambda)L], \quad (27)$$

where $\mu_a^{glucose}(\lambda)$ is the spectral absorption of glucose at the concentration specific to the shot solution, the differential spectral absorption between glucose shot-solution and water (the amount of glucose exceeding water) that is devoid of the source spectral influence is obtained simply as

$$\Delta\mu_a(\lambda) = \mu_a^{glucose}(\lambda) - \mu_a^{water}(\lambda) = \frac{1}{L} ln\left[\frac{T_{water}(\lambda)}{T_{glucose}(\lambda)}\right]. \quad (28)$$

The differential spectral absorption of glucose that exceeds the water as is processed using Eq. (28) from the raw profiles of Fig. 8(A) (after subtracting the dark-counting baseline) is shown in (B). The profile shown in (B) is quite noisy over the 1500-1700 nm region, because of the much lower intensity of the raw signal in (A) in that region. The raw profile shown in (A) carried over the information of the source spectral profile, whereas the differential spectral absorption profile shown in (B) had the source spectral profile compensated. The profile in (B) showed that glucose is spectrally more absorptive than water over the 1100-1300 nm range, which agrees with earlier report [40]. The remaining acquisition of the total diffuse reflectance was thus fixed at the 1100-1300 nm range to maintain the consistency of experimental conditions specific to the parameter configurations of the monochromator-CCD assembly.

## 4.2 Validation of the diffuse reflectance in the CIAD geometry as a function of tissue properties

### 4.2.1. Total diffuse reflectance from a medium of fixed properties as a function of the size of collection

Fig. 9 presents how the total diffuse reflectance from three solid phantoms varies as a function of the size of collection area which was controlled by translating the distance between the focusing lens and the detection fiber over a total displacement of 10 mm. In Fig. 9, the origin of the abscissa (the reference position) corresponded approximately to the minimal size of collection area on the sample surface, determined by "reversing the light path," or equivalently replacing the detector with a temporary light source and observing the resulting focal spot at the sample. Then, using the detector normally again, the fiber tip was displaced in 0.1 mm steps from the reference position in the direction away from the focusing lens, which resulted in the increase of the collection area on the sample, and a change in measured total diffuse reflectance, as shown in Fig. 9. The open-marker curves corresponded to measured data points, the solid lines corresponded to three-dimensional geometric optics simulations performed with Zemax software, and the three broken lines corresponded to simpler two-dimensional geometric optics model calculations as pertinent to Eq. (25). When combined to appear as the three sets of data, they in the descending order correspond to having the absorption coefficients scaled at 1:2:4 but with approximately the same reduced scattering coefficients. Because the minimal spot size of the collection area as estimated by reversing the light path of the collection was not accurate, the reference position of the detector fiber corresponding to the minimal spot size of the collection area was not accurate. However, since the single reference position was used, the displacement from that reference position for measuring from the three solid phantoms was precisely repeated.

The Zemax simulations project the area of the medium surface that can be imaged to the detection fiber. To simulate the light intensity collected by the detection fiber, the profile of the light emitting elements at the sample surface was considered to be non-uniform, Lambertian, and with a diameter of 20 mm. The non-uniformity was configured to exhibit an approximately linear decrease in intensity from the center point. It was employed to approximate the effect of decreasing intensity for light when diffusely scattered farther from the center of the illumination point. Since only the changes in reflectance, and not the absolute reflectance were important, the resulting simulation curves were normalized to the same offsets and peak factors that align the top curve to measurement. To account for the effect of changing the absorption coefficient of the sample, the sample spot light intensity was varied in the simulation. To fit the measured data, the three simulation curves required normalized source light intensities of 1.00, 0.90, and 0.67, resulting in three different simulation curves. In all cases, the normalization factors and offsets were not changed.

The 3D geometric simulations reveal that three mechanisms appear to cause the profiles of roughly three regions as shown in Fig. 9. When near the reference position, the diffusely reflected light is very sharply imaged at the fiber tip plane with a magnification of about M = 0.1. Only diffusely reflected light within a diameter of about 2 mm (= 0.2/M) would ever enter the fiber and reach the detector. That made the total diffuse reflectance to be insensitive to the displacement of the fiber when close to the reference point (which could be better seen if the data points were projected to the negative aspect of the abscissa). The subsequent rise in measured reflectance accompanies the movement of the fiber tip farther from the focusing lens was caused by defocusing. With defocusing, light diffusely reflected from larger diameters on the sample was able to reach the fiber tip, thus coupling greater amounts into the fiber core. The increase of the light intensity over the zone 2 reached a saturation and then decreased. In this case, the continued defocusing increases to such an extent that the diffusely reflected light from the entire surface area becomes too dispersed at the plane of the fiber tip. As the diffusely reflected light spreads beyond the boundary of the fiber core, the coupled intensity stops rising and eventually drops. The third factor that affects the shape of the measured reflectance is the profile of the intensity from the diffusely reflected light. Uniform and Gaussian profiles were also investigated with similar results, but the linear profile produced the closest fit by better matching the experimentally determined displacement at which the reflectance curves peaked. Importantly, all these results support the notion that the CIAD setup is operating as expected. The two-dimensional geometric optics ray calculations also exhibit a similar rising trend in Zone 1, however they do not match the measured behavior at the reference position or the saturation behavior at large displacements. This is not unexpected since, unlike the Zemax simulations, these calculations do not account for several physical realities of the system, including the three-dimensionality of beams, non-uniform diffuse-reflected light intensities, aspheric lens curvatures, and the numerical apertures of the fiber and optical elements imposed by the 1" cage system. An additional cause of the noticeable discrepancies between the 2D geometrical model-prediction and the experimental measurements in the zone marked as (1) starting at the reference position corresponding to the minimal spot sizes was the noise-floor of all sources that may include those of the surface reflection that was impractical to suppress completely even with the implementation of the polarization controllers. The ascending trends as shown in the zone marked as (2) were expectedly consistent among the three sets of assessment. And as the detector-fiber was displaced farther from the zone (2) to reach the zone marked as (3), the total diffuse reflectance degraded as was also revealed by Zemax. The peaking of the three experimental curves of the total diffuse reflectance at the same distance between the collection fiber and the focusing lens suggested to operate the CIAD configuration at the peaking displacement to assure acquiring the maximal amount of signal for otherwise the same configurations.

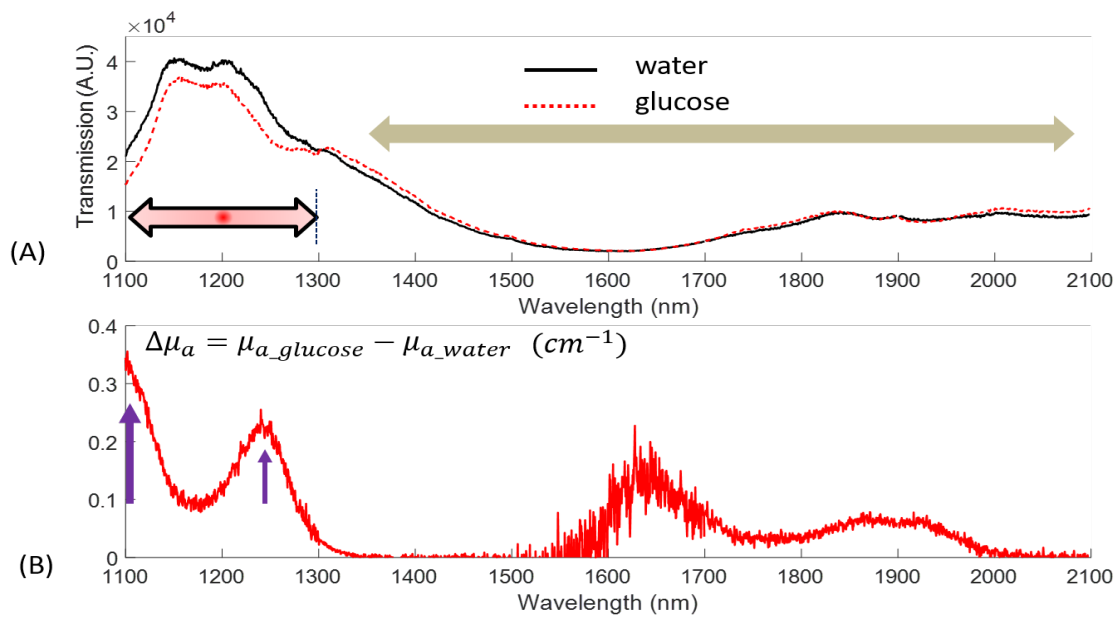

Fig. 8. (A) The raw transmission spectra of the water (solid black line) and the glucose shot solution (dashed red line). (B). The differential spectral absorption of glucose that exceeds the water.

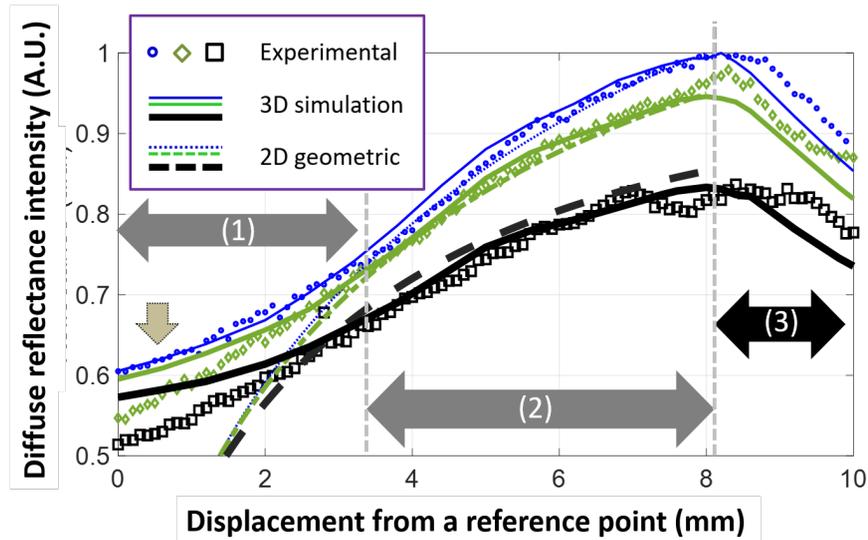

Fig. 9. The total diffuse reflectance from three solid phantoms when assessed with the CIAD geometry at varying size of collection. The size of collection was controlled by translating the distance between the focusing lens and the detection fiber at a step of 100 μm over a total displacement of 10mm. The open-marker data points correspond to the measurements, the solid lines correspond to 3D simulation by using Zemax, and the dotted lines correspond to 2D geometrical modeling.

### 4.2.2. Total diffuse reflectance at a fixed size of detection as a function of the medium scattering

Fig. 10 presents the total diffuse reflectance from an aqueous tissue phantom when assessed with the CIAD geometry configured at a fixed size of collection corresponding to the peaking setting of Fig. 9, as a function of the reduced scattering coefficient of the medium over a range of two orders of magnitude as detailed in 3.3.2. The relative intensity as normalized with respect to that of the one corresponding to the maximal intralipid concentration configured was displayed at linear scale of the abscissa in (A) and logarithmic scale of the abscissa in (B). The circle marks represented experimental measurements, while the solid lines were model estimations. The two large dotted circles in (A) and (B) denote that the reduced scattering properties of the aqueous medium increased from low (coarsely dotted in the circle of (A)) to high (densely dotted in the circle in (B)). The administration of intralipid as shown resulted in a reduced scattering coefficient from 0 to ~18.0 $cm^{-1}$, while the absorption coefficient of the aqueous medium was attributed to be constant at 0.8 $cm^{-1}$ around 1.2μm as it was due primarily to the absorption of water. The predictions by two models are presented, one with the full implementation of Eq. (5) referred to as CIAD model and the other with Eq. (21) denoted by KM model. (A) and (B) combined indicated that the predictions by both the CIAD model and KM model were in excellent agreement with the experimental results as the medium reduced scattering was greater than 10 $cm^{-1}$, whereas the prediction by the CIAD model was considerably more accurate than that by the KM model in the lower scattering region that causes effectively stronger absorption/scattering ratio.

### 4.2.3. Total diffuse reflectance at a fixed size of detection as a function of the medium absorption

Fig.11 presents the total diffuse reflectance from an aqueous tissue phantom when assessed with the CIAD geometry configured at a fixed size of collection corresponding to the peaking setting of Fig. 9, as a function of the absorption coefficient of the medium over a range of one order of magnitude as detailed in 3.3.3. The relative intensity as normalized with respect to that of the one corresponding to the absence of ink injection was displayed at linear scale of the abscissa. The open circle marks represented experimental measurements, while the solid line was the model estimation. The two large shaded circles to the right of the curve plot denote that the absorption property of the aqueous medium increased from low (lightly shaded as the circle 0) to high (darkly shaded as the circle 1). The administration of ink resulted in an absorption coefficient of the medium to increase from 0 to ~19.0 cm$^{-1}$, while the reduced scattering coefficient of the aqueous medium was constant at 18 cm$^{-1}$ around 1.2 µm as it was attributed to the intralipid. The predictions by two models are presented, one with the full implementation of Eq. (5) referred to as CIAD model and the other with Eq. (21) denoted by KM model. The predictions by both models agreed with the experimental results until the measurement ceased to respond to the increase of the ink concentration after approximately 0.7 ml of ink was injected to the total 340 ml of aqueous medium, corresponding to an effective ink concentration of 0.21%.

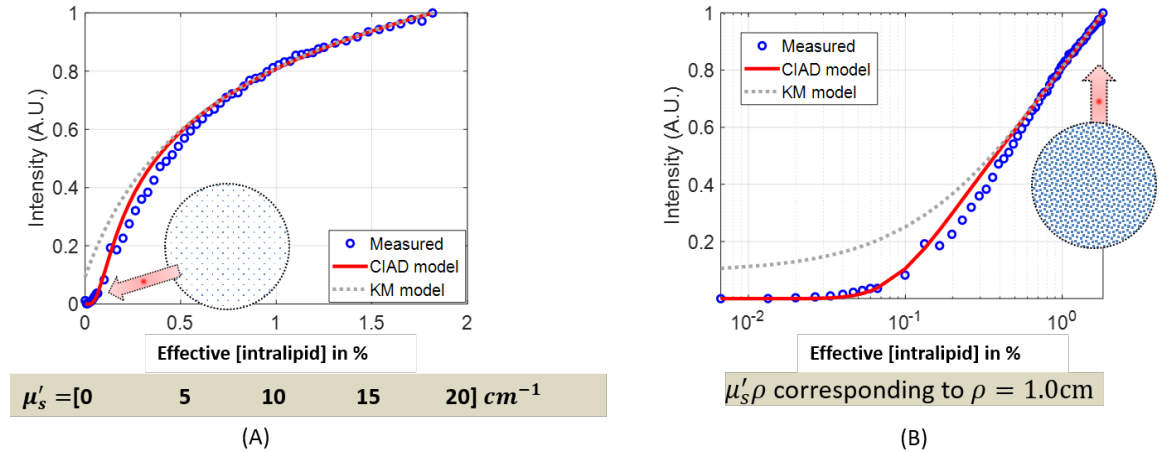

Fig. 10. The total diffuse reflectance from an aqueous tissue phantom when assessed with the CIAD geometry configured at a fixed size of collection corresponding to the peaking setting of Figure. 9, as a function of the reduced scattering coefficient of the medium over a range of two orders of magnitude as detailed in 3.3.2. The relative intensity as normalized with respect to that of the one corresponding to the maximal intralipid concentration configured was displayed at linear scale of the abscissa in (A) and logarithmic scale of the abscissa in (B). The two circles with dots in (A) and (B) denote that the reduced scattering properties of the aqueous medium increased from low (sparsely dotted in the circle of (A)) to high (densely dotted in the circle in (B)). It was indicated that the model-prediction was in excellent agreement with the experimental results as the medium reduced scattering was greater than 10cm$^{-1}$, whereas the model-prediction was considerably off in the lower scattering region.

### 4.3 Assessing glucose dissolved in aqueous tissue medium by diffuse reflectance configured at the CIAD geometry

Fig. 12 presents the total diffuse reflectance from an aqueous turbid medium with the cumulative volume of the glucose control solution varied over three orders of magnitude when assessed at a ~1.1 meters distance, by using the device at the CIAD geometry configured at a fixed size of collection corresponding to the peaking setting of Figure. 9. The diffuse reflectance intensities averaged over 1.1-1.3 µm and normalized with respect to the respective baseline (absent of glucose control solution) were displayed in (A). The four curves of (A) correspond to the same amount of cumulative volume of the glucose control solution dissolved in the aqueous turbid medium. The four curves were however the results of four different glucose-medium conditions. When counted at a descending order of the right-most point of the curve, the first (topmost) one corresponded to the control solution of 0.4% glucose (model 2) dissolved in an aqueous medium containing 2% intralipid, the second (to the topmost) one corresponded to the control solution of 0.2% glucose (model 1) dissolved in an aqueous medium containing 1.14% intralipid, the third (or the second to the lowest) one corresponded to the control solution of 0.2% glucose (model 1) dissolved in an aqueous medium containing 1% intralipid, and the fourth (lowest) one corresponded to the control solution of 0.4% glucose (model 2) dissolved in an aqueous medium containing 1% intralipid.

With the diffuse reflectance processed according to the right-hand side of Eq. (19), the four curves of (A) revealed the grossly linear patterns shown in (B), with respect to the same total volume of control solution dissolved in the aqueous turbid medium. The orders of the four sets of the data points of (B) are opposite to those in (A). Specifically, when counted at a descending order of the right-most point of the curve, the first (topmost) one corresponded to the control solution of 0.4% glucose (model 2) dissolved in an aqueous medium containing 1% intralipid, the second (to the topmost) one corresponded to the control solution of 0.2% glucose (model 1) dissolved in an aqueous medium containing 1% intralipid, the third (or the second to the lowest) one corresponded to the control solution of 0.2% glucose (model 1) dissolved in an aqueous medium containing 1.14% intralipid, and the fourth (lowest) one corresponded to the control solution of 0.4% glucose (model 2) dissolved in an aqueous medium containing 2% intralipid. Linear regression of the four sets of data in (B) resulted in the slopes of respectively 0.0032, 0.0030, 0.0027, and 0.0019, and the $R^2$ values of respectively 0.9872, 0.9902, 0.9919, and 0.9908 at an ascending order of the right-most points.

Notice that the four sets of measurements correspond to the same volume of the control solution injected to the same volume of the aqueous host medium whose scattering properties were adjusted by the amount of intralipid. The volume of the control solution injected was controlled well by using 1ml insulin syringe, whereas the volume of the aqueous host medium was only grossly estimated to be the same as the container had to be refilled after each batch of host medium was exhausted for one set of measurement. In terms of the consistency of measurement, even with magnetic agitation to homogenize the aqueous medium and a data acquisition time of 10 seconds to average over approximately 30 cycles of the agitation, the measurement was still subjective to variations due to the difficulty to grossly control the height and the area of the site of light collection on the curved surface of the aqueous medium yielded to the agitation. These limitations made inter-sample comparison of the absolute intensity of the diffuse reflectance measurements less informative. Whereas, intra-sample evolutions of the relative change of the diffuse reflectance measurements could inform the effect of the only condition of the medium set that has been varied.

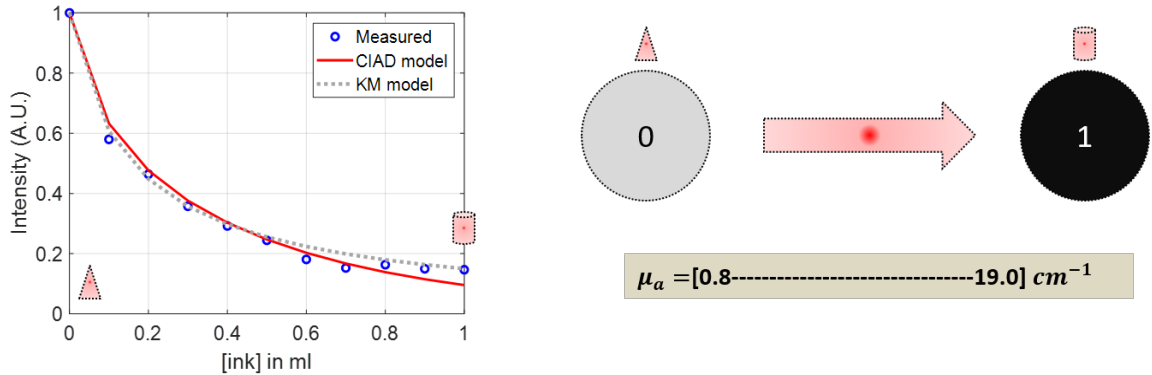

Fig. 11. The total diffuse reflectance from an aqueous tissue phantom when assessed with the CIAD geometry configured at a fixed size of collection corresponding to the peaking setting of Figure. 9, as a function of the absorption coefficient of the medium over a range of one order of magnitude as detailed in 3.3.3. The relative intensity as normalized with respect to that of the one corresponding to the absence of the ink injection was displayed at linear scale of the abscissa. The two circles right to the curve plot denote that the absorption property of the aqueous medium increased from low (lightly shaded as the circle 1) to high (darkly shaded as the circle 1).

Each set of measurements with the linear regression superimposed as shown in Fig. (B) has one of the other three sets to differ in only one controlled condition of the glucose-medium composite. For example, the top-most set marked by the squares and thinner red line differs from the lowest set marked by the diamonds and thicker black line in only the reduced scattering coefficient of the aqueous medium. The lowest set corresponding to an aqueous medium of 2% intralipid and the top-most set corresponding to an aqueous medium of 1% intralipid (making its reduced scattering coefficient to be 50% of the former) were administered with the same set of cumulative volume of the control solution (model 2) having 0.4% glucose. Given that the cumulative volume of the glucose control solution was identical in these two sets, the lowest set had a slope of the change approximately 59% of the top-most set in comparison to an expectation of 50% due to the difference in reduced scattering coefficient. The second lowest set corresponding to an aqueous medium of 1.14% intralipid and the second to the top-most set corresponding to an aqueous medium of 1% intralipid (making its reduced scattering coefficient to be 87.7% of the former) were administered with the same set of cumulative volume of the control solution (model 1) having 0.2% glucose. Given that the cumulative volume of the glucose control solution was identical in these two sets, the lower set had a slope of the change approximately 90% of the upper set in comparison to an expectation of 87.7% due to the difference in the reduced scattering coefficient.

The top-most set and the second to the top-most set were similar in the reduced scattering properties of the aqueous medium corresponding to 1% intralipid and administered with the same cumulative volume of the control solution. The only difference between these two sets was the effective glucose concentration in the control solution injected. Since the effective glucose concentration in the control solution injected was higher for the top-most set than that for the lower set, the slope of the change for the top-most set was expected to be greater than that of the lower set for the same reduced scattering property. However, the slope of the topmost set was only 6.7% greater than that of the lower set, which is substantially smaller than the 100% higher of the effective concentration of the glucose in the control solution in the top-most set than in the lower set. This discrepancy was due to the existence of chromophores other than glucose in control solution to mitigate the changes to diffuse reflectance caused by glucose concentration change. Notice that even though the evolution of the top-most set differed not very remarkably over the conditions assessed in comparison to the set lower to it, the pattern of the difference was consistent. It can be expected that, should the cumulative glucose concentration continue to increase, the absolute difference of the diffuse reflectance between neighboring sets would continue to increase, and that would render stronger the probability of identifying the glucose-origin of the diffuse reflectance.

The terminal effective glucose concentration in the aqueous medium experimented in this study was less than 19 mg/dL. It must be identified that this value does not represent what can be measured for fasting venous plasma glucose levels which are considered normal when ≤100 mg/dL) [41] or the effective interstitial fluid glucose concentration in tissue which is approximately 28 mg/dL lower than the normal levels of fasting venous plasma glucose as the glucose has to be averaged over the capillary and interstitial spaces [42]. A previous approach of optical sensing has shown to be able to identify cutaneous glucose level of approximately 93 mg/dL [23]. Whereas it is the resolution of measurement that will ultimately determine the applicability upon calibration of the baseline. To this end, our preliminary work shows an ability to resolve 8.9 mg/dL difference of the glucose concentration in the aqueous turbid media having the same baseline scattering. This resolution potentiates the applicability, when considering that a CW-photoacoustic contact-based protocol has measured aqueous glucose down to 9.6 mg/dL concentration level with an accuracy of ±3.5 mg/dL at a bias from temperature fluctuations equivalent to a ±11 mg/dL glucose concentration variation [43].

It shall be further clarified that any spectral assessment of glucose in tissue will also be subjected to compromise by chromophores other than water and glucose. To ultimately retrieve the absorption contribution caused by glucose alone, the scattering properties of the tissue shall also be quantified, which would call for techniques based on either diffuse reflectance or other optical methods. The patterns demonstrated in this work suggest that implementing additional spectral discrimination measures will help isolate the effect to the diffuse reflectance intensity caused by glucose absorption. This will require assessing the diffuse reflectance at multiple discrete wavelengths over the spectral range wherein glucose has an absorption contrast over chromophores other than water. An overarching issue also needs to be considered for diffuse reflectance at a distance as great as 1.1 meters as attempted in this work: the need to increase the signal-to-noise ratio. Steady-state measurement as demonstrated in this work was done with an extremely bright broadband light source and intensified camera with high sensitivity. The long projection distance, however, made the signal too weak, thereby sensitive to environmental noise and calibration error. A possible solution is to implement frequency-domain measurements at wavelengths rendering isolation of glucose contribution to diffuse photon remission to take advantage of the coherence detection. Until non-contact diffuse reflectance assessment of glucose at meters distance can be performed at a time-scale of video-rate, practical applications will be difficult.

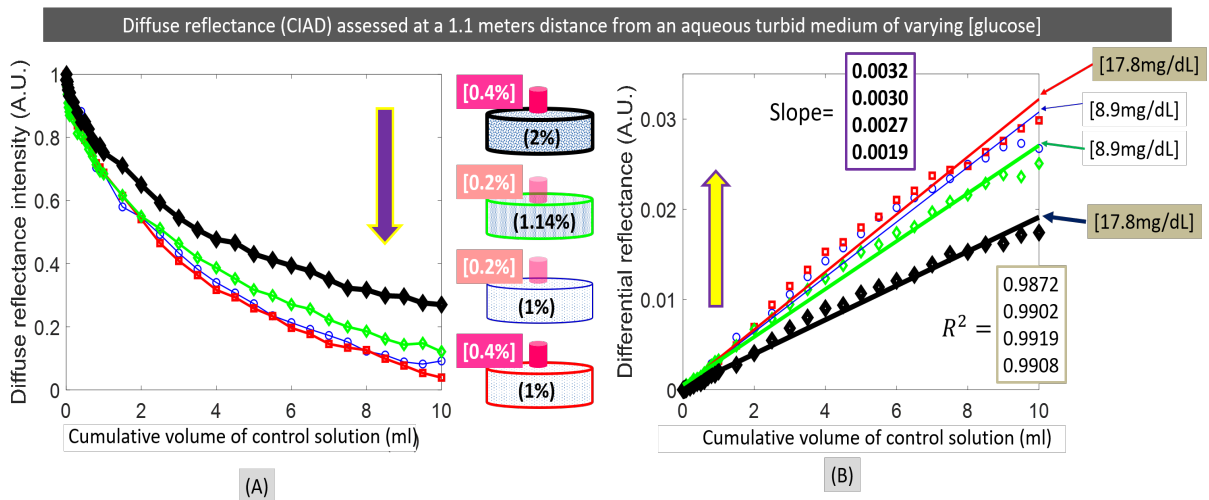

Fig. 12. (A) The total diffuse reflectance from an aqueous turbid medium with the glucose concentration varying over three orders of magnetite when assessed at a 1.1 meters distance. (B) The differential or modified diffuse reflectance processed according to the RHS of Eq. (18) as a function of the cumulative volume of the glucose control solution injected into the host aqueous medium containing intralipid. The four curves in (A) or (B) correspond to four conditions of the glucose-medium contrast as illustrated in the middle vertical panel. The host aqueous medium had intralipid concentration of 1%, 1.14% or 2%. The control solution differed in the effective concentration of glucose, either 0.2% or 0.4%. Each of the curve had one among the three other curves to differ in only one condition of the glucose-medium combination.

## 5. Conclusions

In conclusion, this work has demonstrated that it is feasible to sense glucose in aqueous turbid medium over a distance greater than 1 meter by using diffuse reflectance approaches to probe the absorption/reduced-scattering ratio of the turbid medium. The diffuse reflectance with a working distance of approximately 1.1 meters was developed according to a center-illumination-area-detection (CIAD) geometry that referred to illuminating the sample with a narrow-collimated beam and collecting the diffuse reflectance over the entirety of the surface area rendered by the collection optics and centered on the point-of-illumination. The response of diffuse reflectance has been examined with phantoms by altering independently the size of the collection geometry and the reduced scattering and absorption properties of the medium. When applied to aqueous turbid medium containing glucose control solutions with the cumulative volume varying over three orders of magnitude, a linear relationship expected for the diffuse reflectance over a spectral range of 1.1-1.3 µm and assessed at a 1.1 meters distance differentiated among four conditions of the glucose-medium composition that differed either in the effective glucose concentration up to 17.8mg/dL or the host medium scattering property.